\documentclass[a4paper,11pt]{article}
\pdfoutput=1 

\usepackage{jcappub} 

\usepackage[T1]{fontenc} 
\usepackage{float}
\usepackage{braket}
\usepackage{graphicx}
\usepackage{amsmath}
\usepackage{amsfonts}
\usepackage{amssymb}
\usepackage[english=british]{csquotes}
\usepackage[font=small,labelfont=bf]{caption}

\title{\boldmath An update on adiabatic modes in cosmology and $\delta$N formalism }

 \author{Diego Cruces, }
 \author{Cristiano Germani and }
 \author{Adrian Palomares}

\affiliation{Institut de Ci\`encies del Cosmos (ICCUB) and Universitat de Barcelona, \\
Mart\' i Franqu\`es 1, 08028 Barcelona, Spain }

\emailAdd{dcrucema7@icc.ub.edu}
\emailAdd{germani@icc.ub.edu}
\emailAdd{apalomvi19@alumnes.ub.edu}

\abstract{In this paper, we generalize the Weinberg's procedure to determine the comoving curvature perturbation $\cal R$ to non-attractor inflationary regimes. We show that both modes of $\cal R$ are related to a symmetry of the perturbative equations in the Newtonian gauge.  As a byproduct, we clarify that adiabaticity does not generally imply constancy of $\cal R$, not even in the $k\rightarrow 0$ limit. We then show that there exist non-equivalent definitions of $\delta N$ that would reproduce $\mathcal{R}$ or the uniform density curvature perturbation $\zeta$ at linear order. We have then shown that the perturbative $\delta N$ definition in terms of difference between the number of e-foldings of different gauges, can be extended non-perturbatively at leading order in gradient expansion. Nevertheless, the computer friendly definition in terms of the difference of e-foldings obtained from the evolution of a local FRW Universe, respectively with perturbed and un-perturbed initial conditions, might only give information about the linear order curvature perturbations, contrary to what is stated in the literature.}

\begin{document}
\maketitle
\flushbottom

\section{Introduction}
\label{sec:intro}

The theory of cosmic inflation has become a widely accepted model for the early universe. Inflation is a period of accelerated expansion that was proposed in \cite{guth1981inflationary,starobinsky1980new,sato1981first,linde1982new} to solve some of the problems that  standard cosmology had. Its most impressive achievements were not only to give a reason for the homogeneity and isotropy observed at large scales, but to produce the cosmological perturbations that seed all the large-scale structure of the universe that we see today \cite{mukhanov}. 
The simpler inflationary model, and the one we will consider here, is the single-field scenario in which inflation was driven by a single quantum scalar field, the inflaton.

The unperturbed universe is described by the flat Friedman-Lemaitre-Robertson-Walker (FLRW) metric:
\begin{equation} \label{eq: FLRW metric}
    ds^2=-dt^2+a^2(t)\delta_{ij}dx^idx^j,
\end{equation}
where $a(t)$ is the scale factor. The Hubble parameter is then defined as $\bar{H}(t)\equiv \frac{\dot{a}(t)}{a(t)}$, with the dot serving as a derivative with respect to cosmic time and the bar standing for background quantities. We can also define the number of e-foldings as $N=\int \bar{H} dt$, which is a useful time variable instead of the cosmic time. 

The inflaton-gravity action is 
\begin{equation} \label{eq: unperturbed action}
    S=\frac{1}{2}\int d^4x\sqrt{-g}[M_{pl}^2R-\partial_{\mu}\phi\partial^{\mu}\phi-2V(\phi)] . 
\end{equation}
where $R$ represents the Ricci scalar, $V(\phi)$ the scalar potential, and $M_{pl}$ the reduced Planck mass ($M_{pl}^2=8\pi G_N$).

The metric (\ref{eq: FLRW metric}) is obtained by considering a homogeneous and isotropic universe. Action (\ref{eq: unperturbed action}), with $\bar\phi=\bar\phi(t)$, leads to the following field equations:
\begin{equation}\label{eq: EOM}
    \Ddot{\bar{\phi}}+3\bar{H}\dot{\bar{\phi}}+V'(\bar{\phi})=0,
\end{equation}
\begin{equation}\label{eq: Friedman equation}
    3\bar{H}^2M_{pl}^2=\frac{\dot{\bar{\phi}}}{2}+V(\bar{\phi})\equiv \bar{\rho},
\end{equation}
where $\bar\rho$ stands for the energy density. Eqs (\ref{eq: EOM}-\ref{eq: Friedman equation}) and the metric (\ref{eq: FLRW metric}) determine the evolution of the unperturbed flat universe.

A period of accelerated expansion like inflation, should fulfill: 
\begin{equation} \label{eq: acceleration scale factor}
    \frac{\ddot{a}}{a}=\dot{\bar{H}}+\bar{H}^2=\bar{H}^2\left(1+\frac{\dot{\bar{H}}}{\bar{H}^2}\right)=\frac{\bar{\rho}}{3M_{pl}^2}(1-\epsilon_1)>0
\end{equation} 
in which we defined the first slow-roll (SR) parameter $\epsilon_1\equiv -\frac{\dot{\bar{H}}}{\bar{H}^2}$.

Inflation would be an exactly exponentially expanding (De Sitter) universe, $a(t) = a_0e^{\bar{H}(t-t_0)}$, if $\epsilon_1$ was identically zero. The end of inflation occurs when $\epsilon_1$ gets of order unity. It is then useful to define a parameter that controls the rate of change of $\epsilon_1$, which is the second SR  parameter, $\epsilon_2\equiv \frac{\dot{\epsilon}_1}{\bar{H}\epsilon_1}$. 

Using eqs. (\ref{eq: EOM}-\ref{eq: Friedman equation}), we find the useful equalities:
\begin{equation} \label{eq: SR parameters}
    \epsilon_1\equiv -\frac{\dot{\bar{H}}}{\bar{H}^2}=-\frac{d\ln{\bar{H}}}{dN}=\frac{\dot{\bar{\phi}}^2}{2\bar{H}^2M_{pl}^2};\quad \epsilon_2\equiv \frac{\dot{\epsilon}_1}{\bar{H}\epsilon_1}=\frac{
    d\ln{\epsilon_1}}{dN}=-6-2\frac{V'(\bar{\phi})}{\bar{H}\dot{\bar{\phi}}}+2\epsilon_1.
\end{equation} 

While, for inflation, $\epsilon_1$ is constrained to be small, the second SR parameter is not. Dependently on its value we have the following inflationary regimes:

\begin{itemize}
    \item \textbf{Slow-Roll Inflation (SR), $\epsilon_2>0$}: is an attractor regime in which the field reaches a state of almost constant velocity ($\Ddot{\bar{\phi}}\approx0$).
    \begin{equation}\label{eq: SR eom}
        3\bar{H}\dot{\bar{\phi}}+V'(\bar{\phi})\approx 0.
    \end{equation}
    In this regime, both SR parameters are set to be much smaller than one during inflation and can be written in terms of the derivatives of the potential using the above equation of motion (\ref{eq: SR eom}).
    \begin{equation}\label{eq:Slow-Roll SR parameters}
        \epsilon_1=\frac{M_{pl}^2}{2}\left(\frac{V'(\bar{\phi})}{V(\bar{\phi})}\right)^2;\quad \epsilon_2=2\epsilon_1=M_{pl}^2\left(\frac{V'(\bar{\phi})}{V(\bar{\phi})}\right)^2\Rightarrow \epsilon_1\propto a^0
    \end{equation}
    \item \textbf{Beyond slow-roll, $\epsilon_2<0$}: one particular case that will be treated here is the Ultra-Slow-Roll Inflation (USR). USR is a non-attractor regime in which the field potential is constant $V'(\bar{\phi_0})=V''(\bar{\phi_0})=0$. In this case, 
    \begin{equation}\label{eq: USR eom}
        \ddot{\phi}+3H\dot{\phi}=0\Rightarrow \dot{\phi}\propto a^{-3}
    \end{equation} In this regime, the first SR parameter is set to be much smaller than one, as in the previous case, and it is exponentially decreasing. So that,
    \begin{equation} \label{eq: USR SR parameters}
        \epsilon_2=-6+2\epsilon_1 \Rightarrow \epsilon_1\propto a^{-6} .
    \end{equation}
   Beyond slow-roll models are necessary for primordial black hole formation (PBH) \cite{germani2017primordial, Motohashi:2017kbs}. 
    
\end{itemize}

Even though we only mention the two of the most interesting regimes of inflation to make our points, the results obtained in this paper are applicable to any single-field scenario. 

\section{Linear perturbations} \label{sec: grow in time zeta}
For the scalar sector, the linearly perturbed metric is:
\begin{align} \label{eq: linearized metric}
    ds^2=-(1+2A)dt^2+2a\partial_iBdx^idt+&a^2\left[\left(1+2\psi\right)\delta_{ij}-2\partial_i\partial_jE\right]dx^idx^j \\ \label{eq: definition psi} 
    \psi\equiv D+\frac{1}{3}\delta^{ik}\partial_i\partial_kE&=D+\frac{1}{3}\nabla^2E 
\end{align}
The four functions appearing in the perturbed metric, under the general coordinate transformation 

\begin{equation}\label{eq: gauge transformation}
    \tilde{x}^{\mu}=x^{\mu}+\xi^{\mu} \quad\text{with } \xi_i=\xi^{(V)}_i+\partial_i\xi ,
\end{equation} 
transform as:
\begin{align}\label{eq: linear 1}
    A&\rightarrow \tilde{A}=A-\dot{\xi}^0 \\
    B&\rightarrow \tilde{B}=B-a\dot{\xi}+\frac{\xi^0}{a} \\
    D&\rightarrow \tilde{D}=D-\bar{H}\xi^0-\frac{1}{3}\nabla^2\xi\\
    E&\rightarrow \tilde{E}=E+\xi \\ \label{eq: linear 5}
    \psi&\rightarrow \tilde{\psi}=\psi-\bar{H}\xi^0\ .
\end{align}
The energy-density $ \rho\equiv -T^t_t$ and pressure $p\ \delta^i_j\equiv T^i_j$ where $T^\alpha_\beta$ is the scalar energy-momentum tensor, transform as 
\begin{align}
   \delta \phi&\rightarrow \delta\tilde{\phi}=\delta\phi-\dot{\bar{\phi}}\xi^0 \\ \label{eq: delta rho}
   \delta \rho &\rightarrow \delta\tilde{\rho}=\delta\rho-\dot{\bar{\rho}}\xi^0 \\ \label{eq: linear ulti}
   \delta p &\rightarrow \delta\tilde{p}=\delta p -\dot{\bar{p}}\xi^0.
\end{align}
The gauges that will be central in this paper are:
\begin{itemize}
    \item \textbf{Spatially flat gauge.} The scalar perturbation of the spatial curvature is identically zero. The flat gauge is completely fixed by imposing $\psi_f=D_f=E_f=0$.
    
    \item \textbf{Uniform-density gauge.} This gauge is chosen such that there is no perturbation in the energy density of the field. For the uniform-density gauge one has to impose that $\delta\rho_{ud}=0$ and $E_{ud}=0$.
    
    \item \textbf{Comoving gauge.} The comoving gauge is chosen so that the coordinates follow the flow of the scalar field and hence there is no perturbation of the field. The relevant gauge-fixing condition is  $\delta\phi_c=B_c=0$.
    
    \item \textbf{Newtonian gauge. } It is defined in such a way that the perturbation to the scalar curvature in this gauge, $\psi_N$, equals the Newtonian potential. The gauge-fixing conditions are $B_N=E_N=0$. 
\end{itemize}
We will always use the subindexes $f$, $ud$, $c$ and $N$ to remark the gauge in which every quantity is expressed.

We can define the linearized Mukhanov-Sasaki variable, which is gauge-invariant, as 
\begin{equation}\label{eq: MS variable}
    Q\equiv\delta\phi-\frac{\dot{\Bar{\phi}}}{H}\left(D+\frac{1}{3}\nabla^2E\right)=\delta\phi-\psi\frac{\dot{\Bar{\phi}}}{\bar{H}}.
\end{equation}
This function is related to another useful gauge-invariant variable, which we will refer to as the comoving curvature perturbation $\mathcal{R}$,
\begin{equation}\label{eq: R and Q}
    \mathcal{R}\equiv\psi-\frac{\bar{H}}{\dot{\bar{\phi}}}\delta\phi=-\frac{\bar{H}}{\dot{\bar{\phi}}}Q.
\end{equation}
It is called comoving curvature perturbation as, in comoving gauge, $\mathcal{R}=\psi_c$. Finally, we can also define another gauge-invariant variable $\zeta$:
\begin{equation}
    \zeta\equiv\psi-\frac{\bar{H}}{\dot{\bar{\rho}}}\delta\rho\ ,
\end{equation} 
which is the curvature perturbation in the uniform-density gauge where $\zeta=\psi_{ud}$. The variable $\zeta$ is related to the curvature perturbation $\mathcal{R}$ as we will see in the next section.

\subsection{Linear relation between $\cal R$ and $\zeta$}

It is useful to define the perturbed ``Hubble constant'' $H$ as 
\begin{equation}\label{eq: Hubble parameter}
    H\equiv-\frac{K}{3}\approx\Bar{H}+\delta H=\Bar{H}+\dot{D}-A\Bar{H}-\frac{1}{3}\nabla^2\left(\frac{B}{a}\right).
\end{equation} 
The variation of the action \eqref{eq: unperturbed action} with respect to $A$ leads to the linearized Hamiltonian constraint
\begin{equation} \label{eq: linearized Hamiltonian constraint}
    3\Bar{H}\left(\Bar{H}A-\dot{\psi}\right)+\left(\frac{\nabla^2}{a^2}\right)\left[\Bar{H}\left(a^2\dot{E}+aB\right)+\psi\right]=-\frac{1}{2M_{pl}^2}\left[\dot{\bar{\phi}}\left(\delta\dot{\phi}-\dot{\bar{\phi}}A\right)+V'(\bar{\phi})\delta\phi\right]\ ,
\end{equation} 
while, the variation with respect to $B$, leads to the linearized Momentum constraint,
\begin{equation}\label{eq: linearized Momentum constraint}
    \partial_i\left(\Bar{H}A-\dot{\psi}\right)=\frac{\dot{\bar{\phi}}}{2M_{pl}^2}\partial_i\delta\phi,
\end{equation}
that can be integrated to get:
\begin{equation}\label{momint}
   \Bar{H}A-\dot{\psi}=\frac{\dot{\bar{\phi}}}{2M_{pl}^2}\delta\phi.
\end{equation}

Note that, the integrated version of the momentum constraint has a freedom to include a time-dependent variable. At this point, and without loosing any generality, we are choosing this function to be zero as can be always re-absorbed into the background quantities.

Finally, by combing the Hamiltonian and integrated momentum constraints one finds
\begin{equation}\label{relRZ}
	\zeta=\mathcal{R}-\frac{1}{3\Bar{H}}\Dot{\mathcal{R}}\,.
\end{equation} 
This relation will be very important in the following: Because we are in single-field inflation, there is only one propagating degree of freedom. Thus, the initial conditions of $\zeta$ are tight to those of $\cal R$.

\subsection{Time-evolution of $\cal R$}

In order to find the time dependence of $\cal R$ and, as a consequence of $\zeta$, we need to consider the evolution equation of at least a scalar variable. We consider the perturbed Klein-Gordon equation, this is   
\begin{equation}\label{eq: Linearized KG}
    \delta\Ddot{\bar{\phi}}+3\Bar{H}\delta\dot{\phi}+V''(\bar{\phi})\delta\phi-\frac{\nabla^2}{a^2}\delta\phi+2V'(\bar{\phi})A-\dot{\Bar{\phi}}\left[\dot{A}-3\dot{D}+\nabla^2\left(\frac{B}{a}\right)\right]=0.
\end{equation}
The Mukhanov-Sasaki equation, which is the evolution equation for $Q$, can be written in the following form using eqs. (\ref{eq: linearized Hamiltonian constraint}), (\ref{eq: linearized Momentum constraint}) and (\ref{eq: MS variable}) in (\ref{eq: Linearized KG}), and the definition $\epsilon_3\equiv \dot\epsilon_2/(\epsilon_2 H)$:
\begin{equation} \label{eq: MS equation}
    \ddot{Q}+3\bar{H}\dot{Q}+\left[-\frac{\nabla^2}{a^2}+\bar{H}^2\left(-\frac{3}{2}\epsilon_2+\frac{1}{2}\epsilon_1\epsilon_2-\frac{1}{4}\epsilon_2^2-\frac{1}{2}\epsilon_2\epsilon_3\right)\right]Q=0\,.
\end{equation}
We can then rewrite the MS equation in terms of $\mathcal{R}$ after using the relation (\ref{eq: R and Q}) :
\begin{equation}\label{eq: evolution zeta}
        \ddot{\mathcal{R}}+\bar{H}(3+\epsilon_2)\dot{\mathcal{R}}-\frac{\nabla^2}{a^2}\mathcal{R}=0\,.
\end{equation}
The solution in Fourier modes of eq. (\ref{eq: evolution zeta}) in the long wavelength limit (where the above spatial gradient can be neglected) is
\begin{equation}\label{eq: growing zeta}
    \mathcal{R}_k\simeq C_1^k+C_2^k\int e^{-\int(3+\epsilon_2)Hdt}dt.
\end{equation}  
In SR, $\epsilon_1, \epsilon_2\ll 1$, we then see that the comoving curvature perturbations are roughly constant at super-horizon scales. Things are however different, as it is well known, in non-attractor regimes. For example in USR , $\epsilon_1\ll 1$ and $\epsilon_2\approx-6$, we have $\mathcal{R}\propto a^{+3}$.

The constants $C_1^k$ and $C_2^k$ are determined by the initial conditions and are fixed in the Bunch-Davies vacuum \cite{mukhanov2005physical}. In the long-wavelength limit (super-horizon scales), both solutions are not trivial and have the same k-functional form. We proved this in the appendix \ref{appendix} in agreement with \cite{Romano:2015vxz}.

\section{Adiabatic modes and time evolution of $\mathcal{R}$} \label{sec: Adiabatic modes in cosmology}
It is often stated that in \cite{weinberg2003adiabatic}, Weinberg proved that any adiabatic fluid would generate a roughly constant comoving curvature perturbations at super horizon scales. While it is correct that $\cal R$ always has a constant mode, it was proven in \cite{Romano:2015vxz} that this is not always the dominant one, even in the adiabatic case. This is what we will show in this section by generalizing the procedure that Weinberg used.

(Thermodynamical) adiabatic modes,  are such that \cite{Romano:2015vxz}  
\begin{equation} \label{eq: adiabatic condition}
    \frac{\delta\rho}{\delta p}=\frac{\dot{\bar{\rho}}}{\dot{\bar{p}}}.
\end{equation} 
In the Newtonian gauge, 
\begin{equation}\label{eq: Weinberg proof}
	\dot{\mathcal{R}}=-X+\left(\frac{2\epsilon_1 - (3+\epsilon_2)}{3\bar{H}\epsilon_1}\right)\frac{\nabla^2}{a^2}\psi_N\, .
\end{equation} 

For an adiabatic mode $X=\frac{\dot{\bar{\rho}}\delta p-\dot{\bar{p}}\delta\rho}{3\left(\bar{\rho}+\bar{p}\right)^2}=0$ so that
\begin{equation}
\dot{\cal R}\simeq \left(\frac{2\epsilon_1 - (3+\epsilon_2)}{3\bar{H}\epsilon_1}\right)\frac{\nabla^2}{a^2}\psi_N\, .
\end{equation}
At super-horizon scales, one would be tempted, as in the original Weinberg paper, to assume that $\psi_N$ is a local function and, as a consequence, $\nabla^2\psi_N\sim 0$. This, would lead to a conservation of $\cal R$. However, as we shall shortly see, this is generically incorrect. To prove that, we will consider the case of single-field inflation, specifically in the two cases of SR and USR.

For single-field inflation one has:
\begin{equation}
    \frac{\dot{\bar{\rho}}}{\dot{\bar{p}}}=\frac{3\bar{H}\dot{\bar{\phi}}^2}{3\bar{H}\dot{\bar{\phi}}^2+2V'(\phi)\dot{\bar{\phi}}};\quad \frac{\delta \rho}{\delta p}=\frac{\dot{\bar{\phi}}\delta\dot{\phi}-\dot{\bar{\phi}}^2A+V'(\phi)\delta\phi}{\dot{\bar{\phi}}\delta\dot{\bar{\phi}}-\dot{\phi}^2A-V'(\phi)\delta\phi}\ ,
\end{equation} 
and, for example in specially flat gauge and at leading order in $\epsilon_1$, one gets $\frac{\dot{\bar{\rho}}}{\dot{\bar{p}}}\simeq -1$ and $\frac{\delta\rho}{\delta p}\simeq -1$ for SR and $\frac{\dot{\bar{\rho}}}{\dot{\bar{p}}}= 1$ and $\frac{\delta\rho}{\delta p}= 1$ for USR (note that $X$ is gauge invariant), leading to vanishing non-adiabatic pressure \cite{Romano:2015vxz}. 

Thus, at leading order in slow-roll parameters,
\begin{equation}
	\dot{\cal R}_{SR,USR}\simeq -\left(\frac{ 3+\epsilon_2}{3\bar{H}\epsilon_1}\right)\frac{\nabla^2}{a^2}\psi_N\, .
\end{equation}
From eq. \eqref{eq: growing zeta}, we already know that, although $\dot{\cal R}_{SR}\simeq 0$, ${\cal R}_{USR}$ is growing. Thus, this immediately implies that, at super-horizon scales, $\nabla^2\psi_N$ cannot be gradient suppressed in USR as assumed in the Weinberg analysis.  
 
 \subsection{Time evolution} \label{sec: OUR procedure}

 In Newtonian gauge, the metric can be written as:
\begin{equation}
ds^2=-(1+2A_N)dt^2 + a^2(1+2D_N)\delta_{ij}dx^idx^j\,.
\label{metric_newtonian}
\end{equation}

The scalar set of perturbed Einstein's equations in this gauge is
\begin{itemize}
\item Hamiltonian constraint 
\begin{equation}
    -3\bar{H}^2A_N + 3\bar{H}\dot{D}_N-\frac{\nabla^2}{a^2}D_N = \frac{1}{2M_{pl}^2}\left(\dot{\bar{\phi}}\delta\dot{\phi}_N-\dot{\bar{\phi}}^2A_N +V'(\bar{\phi})\delta\phi_N\right)\,.
    \label{ham_newtonian}
\end{equation}
\item Momentum constraint
\begin{equation}
\partial_i\left(-\bar{H} A_N +\dot{D}_N +\frac{\dot{\bar{\phi}}}{2M_{pl}^2}\delta\phi_N\right)=0\,.
\label{mom_newtonian}
\end{equation}
\item Trace of the spacial Einstein equations 
\begin{equation}
6\dot{\bar{H}}A_N+3\bar{H}\dot{A}_N-3\ddot{D}_N+\frac{\nabla^2}{a^2}A_N + 6\bar{H}^2 A_N -6\bar{H} \dot{D}_N =\frac{1}{M_{pl}^2}\left(2\dot{\bar{\phi}}\delta\dot{\phi}_N-2\dot{\bar{\phi}}^2A_N-V'(\bar{\phi})\delta\phi_N\right)\,.
\label{evext_tr_newtonian}
\end{equation}
\item Traceless part of the spacial Einstein equations
\begin{equation}  \left(\partial_i\partial_j-\frac{1}{3}\delta_{ij}\nabla^2\right)\left(A_N+D_N\right)=0
\label{evext_tl_newtonian}
\end{equation}
\end{itemize}

One can show that the above equations are invariant under the following fields re-definitions
\begin{eqnarray}
	\tilde A_N&=& A_N-\dot\lambda^0\ ,\cr
	\tilde D_N&=& D_{N}-H\lambda^0-\frac{1}{3}\nabla^2 \lambda\ ,\cr
	\delta\tilde\phi&=&\delta \phi-\dot{\bar{\phi}}\lambda^0\ ,
\end{eqnarray}
where
\begin{equation}
	\lambda = -x^2 \frac{f_1(t)}{2} +f_2 (t)\,,
	\label{eta_choice}
\end{equation}
and
\begin{equation}
	\lambda^0(t,\vec{x}) = -a^2(t) x^2 \frac{\dot{f}_1(t)}{2} + f_3 (t)\, .
	\label{lambda_choice}
\end{equation}
As we shall see, the function $f_1(t)$ is going to be of perturbative order. Note that, although the field redefinition contain potentially non-perturbative terms (those proportional to $x^2\equiv \delta_{ij}x^ix^j$) the Einstein's equations do not contain any $x^2$. This term either cancels out or is removed by a Laplacian. Thus, in this sense, the proposed field re-definitions keep the equations at linear order. 

It is interesting to note that, in the UV, those field redefinitions are related to the change of coordinates
\begin{equation}\label{coord_transf}
	t\rightarrow t+\lambda^{0}( t,\vec{x}), \qquad x^i\rightarrow x^i+\partial_i\lambda(t,\vec{ x})\,,
\end{equation} 
which take a Friedmann geometry in Newtonian form \cite{nicolis}. Those coordinate transformations may only be extended to the IR if and only if $\dot f_1=0$. The relation between the coordinate transformation and the symmetry of the Einstein's equations in Newtonian gauge was the starting point of the Weinberg procedure in \cite{weinberg2003adiabatic}.

Both tilded and un-tilded functions solve the {\it same} equations and so are, possibly different, solutions of the {\it same} system. Then, because we are only considering linear differential equations, the difference between the tilded and un-tilded solutions would still represent a solution. Moreover, because this solution is homogeneous, it will be related to perturbations in the long-wavelength limit. 

Different solutions are selected by considering specific boundary and/or initial conditions. The boundary conditions, as we are going to see, are related to the momentum constraint, while the initial conditions to the Bunch-Davies vacuum. The latter will ultimately fix the evolution equation of the variable we would like to consider.

The comoving curvature perturbation $\mathcal{R}$ in Newtonian gauge is
\begin{equation}
	\mathcal{R}= \hat D_N - \frac{\bar{H}}{\dot{\bar{\phi}}}\delta\hat\phi_N\,,
	\label{curv_def_newtonian} 
\end{equation}
where hatted functions are a solution of the linear Einstein equations before imposing any boundary or initial conditions.

We can then consider 
\begin{eqnarray}
	{\cal R}=(\tilde D_N-D_N)-\frac{\bar H}{\dot{\bar\phi}}(\delta\bar\phi_N-\delta\phi_N)=f_1(t)\ .
	\end{eqnarray}
This is simply telling us that, until boundary conditions are imposed, the most generic solution of $\cal R$ is a generic function of time. 

The boundary conditions of our set of differential equations are the integrated version of the momentum constraint. As before, we fix the integrated momentum to zero for the untilded variables
\begin{equation} \label{integrated_mom_newtonian}
	-\bar{H}A_N+\dot{D}_N+\frac{\dot{\bar{\phi}}}{2M_{pl}^2}\delta\phi_N=0\, .
\end{equation}
While the transformation from un-tilded to tilded variables leave invariant the equations of motion, they change the boundary conditions (the integrated momentum constraint) into
\begin{eqnarray} \nonumber
	-\bar{H} \tilde A_N +\dot{\tilde D}_N  +\frac{\dot{\bar{\phi}}}{2M_{pl}^2}\delta\tilde \phi_N - \dot{f}_1(t)=0\ .
	\label{integrated_mom_newtonian2}
\end{eqnarray}
This does not represent a problem as the integration of the momentum constraint precisely leaves the freedom of adding a time-dependent function. 

Combining \eqref{ham_newtonian}, \eqref{evext_tr_newtonian} and the second line of \eqref{integrated_mom_newtonian2} we get the following equation:
	\begin{align} \nonumber
		\ddot{\tilde Q}+3\bar{H}\dot{\tilde Q}+&\left[-\frac{\nabla^2}{a^2}+\bar{H}^2\left(-\frac{3}{2}\epsilon_2+\frac{1}{2}\epsilon_1\epsilon_2-\frac{1}{4}\epsilon_2^2-\frac{1}{2}\epsilon_2\epsilon_3\right)\right]\tilde Q \ +\\
		& + \frac{\dot{\bar{\phi}}}{\bar{H}} \left(\ddot{f}_1(t)+\bar{H}(3+\epsilon_2)\dot{f}_1(t)\right)=0\,.
		\label{MS_equation_mod}
	\end{align}
A solution is such that the tilde MS equation is satisfied, i.e.
\begin{equation}
	\ddot{\tilde Q}+3\bar{H}\dot{\tilde Q}+\left[-\frac{\nabla^2}{a^2}+\bar{H}^2\left(-\frac{3}{2}\epsilon_2+\frac{1}{2}\epsilon_1\epsilon_2-\frac{1}{4}\epsilon_2^2-\frac{1}{2}\epsilon_2\epsilon_3\right)\right]\tilde Q=0\,,
	\label{MS_equation}
\end{equation}
leaving 
\begin{equation}
	\ddot{f}_1(t)+\bar{H}(3+\epsilon_2)\dot{f}_1(t)=0\,.
	\label{curv_equation}
\end{equation}
The solution of this equation is 
\begin{equation}
	f_1(t)=C_1+C_2\int e^{-\int \bar{H} (3+\epsilon_2) dt} dt\, ,
	\label{curv_solution}
\end{equation}
implying that, in Fourier space and in the $k\rightarrow 0$ limit,
\begin{eqnarray}\label{final}
	\mathcal{R}_k=C_1^k+C_2^k\int e^{-\int \bar{H} (3+\epsilon_2) dt} dt\, .
	\end{eqnarray}
Because ${\cal R}_k$ follows a second order differential equation, the solutions \eqref{final} represent the whole set of solutions and the constant $C_i^k$ can now be fixed by initial conditions.

Thus, we have proven that both decaying(growing) and constant modes, are related to a hidden symmetry of the perturbed Einstein equations in Newtonian gauge, extending the analysis of Weinberg.

One may do precisely the same analysis for $\zeta$ instead of $\cal R$ similarly obtaining
\begin{eqnarray}\label{final}
	\zeta_k=c_1^k+c_2^k\int e^{-\int \bar{H} (3+\epsilon_2) dt} dt\, .
\end{eqnarray}
By using the relation \eqref{relRZ} we get
\begin{equation}\label{eq: zeta constants}
	\zeta_k=C_1^{k}+C_2^{k}\int e^{-\int \left(3+\epsilon_2\right)Hdt}dt-\frac{C_2^{k}}{3\Bar{H}}e^{-\int\left(3+\epsilon_2\right)Hdt}\,.
\end{equation} 
In USR, at leading order in $k$, we get
\begin{equation}
	\zeta^{USR}_k=C_1^{k}=c_1^k\,,   
\end{equation} 
while in SR   
\begin{equation}
	\zeta^{SR}_k=C_1^{k}+C_2^{k}\int e^{-\int 3Hdt}dt\,.  
\end{equation}
This is in agreement to the fact that $\zeta$ is non-linearly constant at super-horizon scales \cite{Langlois:2005ii}, up to a decaying mode. 

In the next section we will show how we can relate both $\zeta$ and $\mathcal{R}$ with $\delta N$ via the use of the $\delta N$ formalism formulated in different gauges.
 
\section{Separate Universe approach} \label{sec: SU}

The so-called Separate Universe approach is a particularly simple way to study the evolution of perturbations in the long wavelength limit ($k\rightarrow 0$). This method takes into account perturbations with characteristic wavelength $\lambda$ much larger than a Hubble distance in a local patch $H_p^{-1}$. Under this consideration, we can interpret the region inside the local Hubble patch as an FLRW universe. These conditions are met if the physical wavenumber of the perturbation $\frac{k}{a_p}=\frac{2\pi}{a_p\lambda}$ fulfills $\frac{k}{a_pH_p}\ll 1$ during inflation\footnote{At some point, we will drop the $p$ subscript and we will only call $\frac{k}{aH}$.} \cite{Leach:2001zf,rigopoulos2003separate}. 

Note that, although in this approach any individual patch evolves separately as an unperturbed universe, the ensemble of patch universes evolves as a perturbed universe. 

With respect to the background evolution defined at strictly $k=0$, the local patch evolves, at leading order in $\frac{k}{a \bar H}\rightarrow 0$ as
\begin{align} \label{eq: patch hubble parameter}
    &H_p\approx \Bar{H}+\delta H=\Bar{H}+\dot{D}-A\Bar{H}-\frac{1}{3}\nabla^2\left(\frac{B}{a}\right) \\ \label{eq: patch scalar field} &\phi_p\approx\Bar{\phi}+\delta\phi \\ \label{eq: energy-density patch} &\rho_p=\frac{1}{2}\left(\frac{d\phi_p}{dt_p}\right)^2+V(\phi_p)\approx\frac{\dot{\Bar{\phi}}^2}{2}+V(\Bar{\phi})+\dot{\Bar{\phi}}\delta\dot{\phi}-\dot{\Bar{\phi}}^2A+V'(\Bar{\phi})\delta\phi \\   \label{eq: patch time}
     &dt_p\approx(1+A)dt\ ,
\end{align} 
where no gauge fixing has been used.

The term $\nabla^2 B$, that we have kept, it is going to give a contribution to the evolution at zeroth order in long wavelength \cite{cruces2022review}. In other words, generically, $B$ is a {\it non-local} function, as it can be more easily appreciated in the flat gauge: by using eq.(\ref{eq: linearized Hamiltonian constraint}) and eq. (\ref{eq: linearized Momentum constraint}), one finds
\begin{equation} \label{eq: nabla B USR}
    \nabla^2\left(\frac{B_f}{a}\right)=\Bar{H}\left(A_f\frac{\epsilon_2}{2}-\frac{\epsilon_1}{\dot{\bar{\phi}}}\delta\dot{\phi}_f\right)=\bar{H}A_f\left(\frac{\epsilon_2}{2}-\frac{\delta \dot{\phi}_f}{H\delta\phi_f}\right)\ ,
\end{equation} 
showing that $B_f$ is indeed non-local and $\nabla^2 B_f$, i.e., it is not suppressed at super-horizon scales.

Specifically, in SR (see e.g. \cite{pattison2019stochastic} and appendix B of \cite{cruces2022stochastic})
\begin{equation*}
    \left(\frac{\delta\dot{\phi}_f}{H\delta\phi_f}\right)_{\textbf{SR}}=\frac{\epsilon_2}{2}+\mathcal{O}\left((k/aH)^2\right)\ ,
\end{equation*} 
and $\nabla^2B$ is negligibly small in SR. 

On the contrary, as in USR $\epsilon_2\approx -6+\mathcal{O}(\epsilon_1)$, we have
\begin{equation*}
    \left(\frac{\delta\dot{\phi}_f}{\bar{H}\delta\phi_f}\right)_{\textbf{USR}}\sim\mathcal{O}(\epsilon_1)\, .
\end{equation*}
Therefore, in USR, $B$ is not super-horizon suppressed while decaying as $\nabla^2 B/a \propto a^{-3}$. Although this would nevertheless seem negligible, we have already checked in section \ref{sec: Adiabatic modes in cosmology} that it can play a fundamental role when multiplying by a growing function so, in order to be cautious, we will keep terms like $\nabla^2 B/a$ in the following.

\section{$\delta$N formalism} \label{sec: delta N}
 The $\delta N$ formalism is a tool used for computing the evolution of cosmological perturbations at super-horizon scales \cite{Starobinsky:1982ee,Salopek:1990jq,Sasaki:1995aw,Lyth:2004gb,Lyth:2005fi, abolhasani2019deltan,sugiyama2013delta}.
 
The number of e-folds in a given super-horizon patch is 

 \begin{equation}
     N_p=\int_{t_p(\bar{t}^0,\mathbf{x})}^{t_p(\bar{t}^e,\mathbf{x})}H_p(t_p)dt_p\simeq\int_{\bar{t}^0}^{\bar{t}^e}H_p(\bar{t})\alpha_p(\bar{t})d\bar{t}\,,
     \label{N_definition}
 \end{equation} 
where $\alpha_p$ is the lapse function of each super-horizon patch defined as $dt_p=\alpha_p d\bar{t}$ and we have expanded everything at leading order in gradient expansion. Note that this definition of $N$ is not gauge invariant as it depends on the specific gauge relating the local to the background coordinates. 

If we now perturb \eqref{N_definition} with respect to a FLRW background we get the expression for the number of e-folds in a perturbed universe

\begin{equation}
N_p \simeq \int_{\bar{t}^0}^{\bar{t}^e}\left(\bar{H}+\dot{D}-\frac{\nabla^2}{3}\left(\frac{B}{a}\right)\right)d\bar{t}\,,
\label{N_definition_pert}
\end{equation}
where we have not fixed any gauge yet. A gauge transformation \eqref{eq: gauge transformation}-\eqref{eq: linear ulti} would lead to
\begin{equation}
\int_{\bar{t}^0}^{\bar{t}^e}\left(\bar{H}+\dot{D}-\frac{\nabla^2}{3}\left(\frac{B}{a}\right)\right)d\bar{t} \longrightarrow \int_{\bar{t}^0}^{\bar{t}^e}\left(\bar{H}+\frac{d}{d\bar{t}}\left(D-\bar{H}\xi^0\right)-\frac{\nabla^2}{3}\left(\frac{B}{a}\right)-\frac{\nabla^2}{3}\left(\frac{\xi^0}{a^2}\right)\right)d\bar{t}\,.
\label{N_transf}
\end{equation}
 We will choose gauges where $\nabla^2\xi^0$ is next to leading order in gradient expansion\footnote{This is not possible in Newtonian gauge as we checked in section \ref{sec: Adiabatic modes in cosmology}}. With this in mind, the number of e-folds transform as:

\begin{equation}
N_p \rightarrow N_p - \bar{H}\xi^0\Big|_{\bar{t}^e}+\bar{H}\xi^0\Big|_{\bar{t}^0}
\end{equation}
 
If $\bar{H}\xi^0\Big|_{\bar{t}^e}$ and $\bar{H}\xi^0\Big|_{\bar{t}^0}$ represent different gauges, let us say gauge $\mathcal{A}$ and $\mathcal{B}$, we will say that we have chosen an interpolating gauge between $\mathcal{A}$ and $\mathcal{B}$ and we will write $N_{\mathcal{A}}^{\mathcal{B}}$.  

We then finally define $\delta N$ as follows
\begin{equation}
\delta N \equiv N_{\mathcal{A}}^{\mathcal{B}}-N_{\mathcal{A}'}^{\mathcal{B}'}\,.
\end{equation} 
In the following we are going to see that $\zeta$ and $\cal R$ correspond to specific (and different) choices of gauges.

\begin{itemize}
\item The curvature perturbation at uniform density is obtained by choosing ${\mathcal{A}}$ and ${\mathcal{B}}$ respectively in flat and uniform density gauges while ${\mathcal{A}'} $ and ${\mathcal{B}'}$ in flat gauge. 

Indeed, with this choice
\begin{equation}
\delta N_{\zeta}=N_{f}^{ud}-N_{f}^{f} = \left(D-\bar{H}\xi^0_{ud}\right)\Big|_{\bar{t}^e}\,.
\end{equation}
Now because in the uniform density we have $(\delta\rho - \dot{\bar{\rho}}\xi_{ud}^0)\Big|_{{\bar t}^e}=0$, we get
\begin{equation}
\delta N_{\zeta}=N_{f}^{ud}-N_{f}^{f}=\left(D-\frac{\bar{H}}{\dot{\bar{\rho}}}\delta\rho\right)\Bigg|_{\bar{t}^{e}}=\zeta (\bar{t}^e)\,,
\label{deltaN_zeta}
\end{equation}
where the last equality comes by fixing $\xi$ such that $E=0$. 

\item The comoving curvature perturbation is instead obtained by taking $\mathcal{A}$ and ${\mathcal{B}}$ to be flat and comoving gauges, while $\mathcal{A}'$ and $\mathcal{B}'$ are flat gauges. 

Then we get, similarly as before,
\begin{equation}
\delta N_{\mathcal{R}}=N_{f}^{c}-N_{f}^{f}=\left(D-\frac{\bar{H}}{\dot{\bar{\phi}}}\delta\phi\right)\Bigg|_{\bar{t}^{e}}=\mathcal{R} (\bar{t}^e)\,,
\label{deltaN_R}
\end{equation}
\end{itemize}
 As we have already discussed, non-local terms generically make $\zeta\neq{\cal R}$. Ignoring those terms (see for example \cite{Dias:2015rca,Dias:2014msa}), would incorrectly imply $\zeta={\cal R}$ and, in turn, $\delta N_{\zeta}=\delta N_{\mathcal{R}}$. While in SR the difference between $\zeta$ and $\cal R$ is negligible (it is proportional to a decaying mode) in other regimes, like in USR, the difference is exponentially large.

The formulation of the $\delta N$ formalism presented in this section can be generalized to include non-linear effects. In the 3+1 (or ADM) formulation of general relativity, the metric takes the following form

\begin{equation}
ds^2 = -\alpha^2 dt^2 + \gamma_{ij}\left(dx^i + \beta^i dt \right)\left(dx^j+\beta^j dt\right)\,.
\label{metric_ADM}
\end{equation}
where $\alpha$ is the lapse function, $\beta^i$ is the shift vector, and the spatial metric $\gamma_{ij}$ can be redefined as $\gamma_{ij}=a(t)^2e^{2D^{NL}}\tilde{\gamma}_{ij}$, with $\det (\tilde{\gamma}_{ij})=1$. The scale factor $a(t)$ is chosen to have the same functional form and initial conditions of the background. 

At leading order in gradient expansion, we can define the number of e-folds \eqref{N_definition} with the trace of the extrinsic curvature of the spatial metric $K$ as follows

 \begin{equation}
     N(t^0,t^e,x^i)\equiv -\frac{1}{3}\int_{t^0}^{t^e}K(t,x^i)\alpha(t,x^i)\Big|_{x^i=\text{constant}}dt\,,
     \label{N_definition_NL}
 \end{equation} 
where $K$ can be written in terms of the variables in \eqref{metric_ADM}

\begin{equation}
K=-3\frac{H}{\alpha}-3\frac{\dot{D}^{NL}}{\alpha} + \frac{D_i\beta^i}{\alpha}\,,
\label{extrinsic_K_def} 
\end{equation}
where $D_i$ is the covariant derivative with respect to the spatial metric $\gamma_{ij}$. We can then write the number of efolds in \eqref{N_definition_NL} as follows:

\begin{equation}
N(t^0,t^e, x^i)=\int_{t^0}^{t^e}\left(H + \dot{D}^{NL} - \frac{D_i\beta^i}{3}\right)d t\,,
\end{equation}
which is the non-linear version of expression \eqref{N_definition_pert}. We note that, even if it is not generically of higher order in gradient expansion \cite{sugiyama2013delta,cruces2022review,cruces2022stochastic}, $D_i \beta^i$ always decays as the inverse volume, as we have shown in \eqref{eq: nabla B USR}. Neglecting this term, the number of e-folds takes a very simple form

\begin{equation}
N(t^0,t^e, x^i)=\bar{N}(t^0,t^e)+D^{NL}\left(t^e,x^i\right)-D^{NL}\left(t^0,x^i\right)\,.
\end{equation}

As in linear case, we can choose a gauge transformation that interpolates different hypersurfaces for the initial and final times. If we now define a non-linear version of the $\delta N$ formalism (let us call it $\Delta N$) in the same way as we did in \eqref{deltaN_zeta} and \eqref{deltaN_R} we get

\begin{equation}
\Delta N_{\zeta}=D_{ud}^{NL}(t^e,x^i)\,,
\end{equation}

\begin{equation}
\Delta N_{\cal R}=D_{c}^{NL}(t^e,x^i)\,.
\end{equation} 
Whether $D_{ud}^{NL}$ and $D_{c}^{NL}$ are respectively the non-linear generalization of the uniform density curvature perturbation $\zeta$ and of the comoving curvature perturbation $\mathcal{R}$, is however still though an open problem.
 
\subsection{$\delta N$ formalism in terms of inflation initial conditions}

The $\delta N$ described in the previous section is really of little practical use. The reason is that in order to get the $\Delta N$, one would need to know already the solution of the locally perturbed metric in terms of background coordinates. 

By using the separate universe approach more seriously, however, one might hope to obtain the curvature perturbations just by solving the evolution equations in a local FRW universe and subtract the number of e-folds of the background \cite{Dias:2015rca,Dias:2014msa,Garriga:2015tea,Matsuda:2009kp,Vennin:2015hra,Matarrese:2018qqo,Abolhasani:2019cqw,Suyama:2012wi}. The difference between the evolution of one patch to another would then be related to different initial conditions necessary to solve the scalar evolution. Although linearly this idea is extremely powerful, as we are going to see non-linearly would not help more than what discussed in the previous section.

As before, fully non-linearly but at super-horizon scales we can define
\begin{equation}
	\Delta N= -\frac{1}{3}\int_{t^0}^{t^e}K\left(t,x^i\right)\alpha\left(t,x^i\right)\Bigg|_{x^i=constant} dt - \int_{t^0}^{t^e}H\left(t\right)d t\,.
	\label{DeltaN_zeta}
\end{equation}

At leading order in gradient expansion, the continuity equation takes the following form:
\begin{equation}
	\frac{1}{\alpha}\frac{d\rho}{d t}=K\left(\rho+P\right)\,,
	\label{continuity_eqn}
\end{equation}
which means that we can write $\Delta N$ in \eqref{DeltaN_zeta} as follows

\begin{equation}
	\Delta N=-\frac{1}{3}\int_{t^0}^{t^e}\frac{\dot\rho}{\rho+P}dt-\int_{t^0}^{t^e}H(t)dt\,,
	\label{DeltaN_zeta2}
\end{equation}
where $\rho$ and $P$ are respectively the energy density and pressure associated to the inflaton in the local patch.

The idea is now to change the integration variable from $t$ to $\rho$. Obviously, this can only be done if the pressure is adiabatic, i.e., $P=P(\rho)$.

At leading order in gradient expansion
\begin{equation}
	\rho=\frac{1}{2\alpha^2}\left(\frac{d\phi}{dt}\right)^2+V\left(\phi\right)\,, \qquad P=\frac{1}{2\alpha^2}\left(\frac{d\phi}{dt}\right)^2-V\left(\phi\right)\,,
\end{equation}
where we are neglecting $D_i\beta^i$ because, as justified before, it always decays as the inverse volume.

Thus, generically  
	\begin{equation}
		P+\rho= \frac{1}{\alpha^2}\left(\frac{d\phi}{d\bar{t}}\right)^2=\frac{2}{3}\epsilon_1\rho\,,
	\end{equation}
	where we have used the Hamiltonian constraint at leading order in gradient expansion 
	\begin{equation}
		M_{PL}\alpha^2 K^2=3\rho
	\end{equation} 
	and defined 	
	\begin{equation}
		\epsilon_1\equiv\frac{9}{2 K^2\alpha^4 M_{PL}^2}\left(\frac{d\phi}{dt}\right)^2\,.
	\end{equation}
Generically, there is no hope that $\epsilon_1$ is a function of $\rho$. In other words, even if in gradient expansion everything is only function of $t$, it is not guaranteed that $\rho$ is a monotonic function of time. Thus, one may not be able to use the energy density as a time variable.
	
In the perturbative regime, order by order, the adiabaticity condition $P=P(\rho)$ might be however approximately satisfied. For example, at linear level in SR, $\epsilon_1$ is roughly constant at the background level and thus $P+\rho\propto\rho$ (we would like to stress that $\epsilon_1$ is {\it not} necessarily constant non-perturbatively). USR, is instead a special case as $V(\phi)=\Lambda$ and thus $\rho+P=2\left(\rho-\Lambda\right)$ always.  
	
In the regime in which the adiabaticity condition is satisfied, we could, as in previous section, chose a gauge interpolating the flat to uniform density slicing and write 
\begin{equation}
	\Delta N_{\zeta}=-\frac{1}{3}\int_{\rho^0_f}^{\bar{\rho}^e}\frac{d\rho}{\rho+P}-\int_{t^0}^{t^e}H(t)dt\,.
	\label{DeltaN_zeta3}
\end{equation}
Even in the adiabatic case however, the practical use of this formula is only in the perturbative regime. The reason is that $\rho_f^0$, and its probability distribution, might only be obtained by perturbation theory. Moreover, beyond linearity, what $\Delta N_{\zeta}$ is observationally related to is still under debate, althogh we might be tempted to define the non-linear $\zeta$ as $\Delta N_{\zeta}$.

At linear order instead everything simplifies. For example in USR
\begin{equation}
	\delta N_{\zeta}=-\frac{1}{3}\int_{\bar{\rho}^0+\delta\rho_f^0}^{\bar{\rho}^e}\frac{d\rho}{2(\rho-V)}+\frac{1}{3}\int_{\bar{\rho}^0}^{\bar{\rho}^e}\frac{d\bar{\rho}}{2(\bar{\rho}-V)}=\frac{\delta\rho^0_f}{6\left(\bar{\rho}-V\right)}=-\frac{\bar{H}}{\dot{\bar{\rho}}}\delta\rho^0_f=\zeta(\bar{t}^0)\,,
	\label{DeltaN_zeta_USR}
\end{equation}
where $\zeta(\bar{t}^0)$ is the uniform density curvature perturbation at $t=\bar{t}^0$ and linearity in $\delta\rho^0_f$ has been used. 

Note that the only reason why we are recovering $\zeta(\bar{t}^0)$ in \eqref{DeltaN_zeta_USR} and not $\zeta \left(\bar{t}^e\right)$, is because the integrability of \eqref{DeltaN_zeta2} in terms of $\rho$ actually implies a conservation of $\zeta$ at super-horizon scales and hence $\zeta(\bar{t}^0)=\zeta(\bar{t}^e)$. Conversely, if $P \neq P(\rho)$, perturbations are not adiabatic anymore and $\zeta$ is not conserved at superhorizon scales. 

In the case of $\Delta N_{\cal R}$, all the discussions above for $\Delta N_\zeta$ still hold, the only difference is that one needs now to achieve integrability in the $\phi$ variable.

\section{Conclusions}
In this paper, we extended the study of Weinberg  \cite{weinberg2003adiabatic} to non-attractor models of inflation by finding a new non-local symmetry of the perturbative Einstein equations in Newtonian gauge. We then showed that both the constant and the growing (or decaying for attractor models of inflation) modes of the comoving curvature perturbation $\cal R$, are related to this symmetry. Restricting this symmetry to be local, as in the original Weinberg paper, would imply to select the constant mode. As a byproduct, one can then show that for general adiabatic fields, the comoving curvature perturbation is not necessarily constant and differs, by the growing mode, to the uniform density curvature perturbation $\zeta$.  

 This result has an immediate repercussion in the so-called $\delta N$ formalism. Because the number of e-folds is defined as an integral between some initial and final hypersurface, we reviewed here that the $\delta N$ formalism requires a gauge fixing in both hypersurfaces. Depending on the gauges chosen, one can relate the $\delta N$ formalism to correlators of $\zeta$ or $\mathcal{R}$. More concretely, correlators of $\zeta$ involve the $\delta N$ from the flat to the uniform density gauge while $\mathcal{R}$ needs the $\delta N$ to be formulated from flat to comoving gauge. These gauges are not generically interchangeable. We have also shown that the linear version of the $\delta N$ formalism when non-local terms are included still gives the same result as the linear formulation. Nevertheless, the interpretation of the $\delta N$ formalism as the e-folds difference between a FRW evolution with perturbed initial condition and the background e-folds, is only valid in the perturbative regime.

\acknowledgments

AP would like to thank David Seery for discussions. DC would like to thank Misao Sasaki and Vincent Vennin for comments on the first version of this paper. CG and DC are parcially supported by by the Unidad
de Excelencia Maria de Maeztu Grant No. CEX2019-
000918-M, and the Spanish national grants PID2019-
105614GB-C22, PID2019-106515GB-I00. DC is supported by the Spanish MECD fellowship PRE2018-086135.

\appendix
\section{$k$-dependence of the two modes of $\cal R$}\label{appendix}

In this appendix we will compute the $k$-dependence of the two modes of $\mathcal{R}$ by using the Bunch-Davies vacuum as initial condition \cite{Bunch:1978yq}.

In conformal time ($d\tau=\frac{dt}{a}$), the solution for the Mukhanov-Sasaki variable $Q$ in the $k\rightarrow 0$ limit with Bunch-Davies initial conditions, we have

\begin{equation}
Q_{\mathbf{k}} \simeq -i \frac{e^{\frac{i}{2}\left(\nu +\frac{1}{2}\right)}2^{\nu +\frac{1}{2}}}{a\sqrt{\pi}}\sqrt{-\tau}\left(-k\tau\right)^{-\nu}\left( \Gamma \left[\nu\right] - \frac{1}{4} \Gamma \left[\nu-1\right] \left(-k\tau \right)^2\right)\,,
\label{eq:Qsolution}
\end{equation}

where

\begin{equation}
    \nu^2=\frac{1}{4} + \left(\tau \mathcal{H}\right)^2\left(2-\epsilon_1+\frac{3}{2}\epsilon_2+\frac{1}{4}\epsilon_2^2 - \frac{1}{2}\epsilon_1\epsilon_2+\frac{1}{2}\epsilon_2\epsilon_3\right)\,,
\end{equation}
is a constant\footnote{Note that \eqref{eq:Qsolution} requires $\nu$ to be a constant. This means that for USR, where $\epsilon_1 \sim a^{-6}$, we must use $\nu^2=\frac{1}{4} + 2\left(\tau \mathcal{H}\right)^2$}. Here we have defined $\mathcal{H}=\frac{a'}{a}$,  where a prime denotes a derivative with respect to conformal time.

We can now compute $\frac{Q'_{\mathbf{k}}}{\mathcal{H}Q_{\mathbf{k}}}$. The result is:

\begin{equation}
    \frac{Q'_{\mathbf{k}}}{\mathcal{H}Q_{\mathbf{k}}} \simeq \frac{1-2\nu - 2\mathcal{H}\tau}{2\mathcal{H}\tau}+\frac{1}{2\mathcal{H}\tau(\nu-1)}\left(-k\tau\right)^2
\end{equation}

Finally, we can compare the derivative of the long wavelength solution of $\mathcal{R}_{\mathbf{k}}$ i.e $\mathcal{R}_{\mathbf{k}}'= \frac{C_2^k}{a^2\epsilon_1}$ with 

\begin{align} \nonumber
        \mathcal{R}_{\mathbf{k}}'&\equiv\frac{\mathcal{H}^2 Q}{\phi '}\left(\frac{\epsilon_2}{2}-\frac{Q_{\mathbf{k}}'}{\mathcal{H}Q_{\mathbf{k}}}\right)\\ \nonumber
        & \simeq \frac{\phi'}{2M_{PL}^2\epsilon_1}\Bigg\{ -i \frac{e^{\frac{i}{2}\left(\nu +\frac{1}{2}\right)}2^{\nu +\frac{1}{2}}}{a\sqrt{\pi}}\sqrt{-\tau}\left(-k\tau\right)^{-\nu}\left[\left(\frac{\epsilon_2}{2}-\frac{1-2\nu - 2\mathcal{H}\tau}{2\mathcal{H}\tau}\right)\Gamma[\nu]\right. \\
        &+ \left.\left(\frac{1}{4}\left(\frac{1-2\nu - 2\mathcal{H}\tau}{2\mathcal{H}\tau}-\frac{\epsilon_2}{2}\right)\Gamma[\nu-1]-\frac{\Gamma[\nu]}{2\mathcal{H}\tau(\nu-1)}\right)\left(-k\tau\right)^2\right]\Bigg\}\,,
\end{align}
which gives:

\begin{align} \nonumber
        C_2^k & \simeq \frac{a^2\phi'}{2M_{PL}^2}\left( -i \frac{e^{\frac{i}{2}\left(\nu +\frac{1}{2}\right)}2^{\nu +\frac{1}{2}}}{a\sqrt{\pi}}\sqrt{-\tau}\left(-k\tau\right)^{-\nu}\right)\left[\left(\frac{\epsilon_2}{2}-\frac{1-2\nu - 2\mathcal{H}\tau}{2\mathcal{H}\tau}\right)\Gamma[\nu]\right. \\
        &+ \left.\left(\frac{1}{4}\left(\frac{1-2\nu - 2\mathcal{H}\tau}{2\mathcal{H}\tau}-\frac{\epsilon_2}{2}\right)\Gamma[\nu-1]-\frac{\Gamma[\nu]}{2\mathcal{H}\tau(\nu-1)}\right)\left(-k\tau\right)^2\right]\,,
\end{align}

\begin{itemize}
    \item For Slow-Roll we can write $\tau^{SR}= -\frac{1}{\mathcal{H}}\left(1+\epsilon_1+\mathcal{O}\left(\epsilon_1^2\right)\right)$ and $\nu^{SR}=\frac{3}{2}+\epsilon_1+\frac{\epsilon_2}{2} + \mathcal{O}\left(\epsilon_1^2\right)$ so, neglecting $\mathcal{O}(\epsilon_1^2)$ terms we get:
    \begin{equation}
        \left(\frac{1-2\nu - 2\mathcal{H}\tau}{2\mathcal{H}\tau}-\frac{\epsilon_2}{2}\right) \sim \epsilon_1^2\,,
    \end{equation}
    which means that $ C_2^k \sim \frac{\epsilon_1^2}{k^{3/2}}$ for SR. If one neglects $\mathcal{O}\left(\epsilon_1^2\right)$ terms, then $ C_2^k \sim k^{1/2}$ and it can be set to zero in the $k\rightarrow 0$ limit. On the other hand, we can see that for SR backgrounds $C_1^k \sim k^{-3/2}$.

    \item For Ultra-Slow-Roll we have $\tau^{SR}= -\frac{1}{\mathcal{H}}\left(1+\mathcal{O}\left(\epsilon_1\right)\right)$, $\nu^{SR}=\frac{3}{2}+\mathcal{O}\left(\epsilon_1\right)$ and $\epsilon_2 = -6+\mathcal{O}\left(\epsilon_1\right)$ so

\begin{equation}
     C_2^k = \frac{1}{k^{3/2}}\left(a \phi' \mathcal{H}\right)\left(-i\frac{e^i}{M_{PL}^2}\right)\,.
\end{equation}
As it can be seen from the background equations, the first parenthesis is a constant up to $\mathcal{O}(\epsilon_1)$. It is then clear that $ C_2^{k} \sim k^{-3/2}$ in this case. On the other hand,  $ C_1^k \sim \frac{\epsilon_1^0} {k^{3/2}}$ for USR backgrounds which, as we stated in the main text, has the same k-dependence as the growing mode and it can only be set to zero if we consider $\epsilon_1^0=0$.

\end{itemize}

\end{document}